\documentstyle[epsfig,aps,prl,twocolumn]{revtex}
\newcommand{\etal}{{\em{et al.}}}

\newcommand{\mW}{m_W}

\newcommand {\Be}{{\cal{B}}_{e}}
\newcommand {\Bm}{{\cal{B}}_{\mu}}

\newcommand {\Bl}{{\cal{B}}_{\ell}}

\newcommand {\MTval}     {(1776.96^{+0.18+0.25}_{-0.21-0.17})\mathrm{MeV}}   
\newcommand {\TAUTval}   {(290.55     \pm 1.06)        \mathrm{fs}}    
\newcommand {\BRTEval}   {(17.786     \pm 0.072)       \%}             
\newcommand {\BRTMval}   {(17.356     \pm 0.064)       \%}             
\newcommand {\GF}        {G_{\mathrm{F}}}

\newcommand {\GFMUval}   {(1.16639  \pm 0.00002) \times 10^{-5}   \mathrm{GeV^{-2}}} 
\newcommand {\DELTAEval} { -0.0008 \pm 0.0055} 
\newcommand {\DELTAMval} { +0.0026 \pm 0.0053}
\newcommand {\ETAval}     {0.009 \pm 0.022}     
\newcommand {\KAPPAMval}  {0.001 \pm 0.008}     
\newcommand {\KAPPAEval}  {0.00  \pm 0.16}     %

\newcommand {\ETAlim}     {-0.034 < {\eta_{\tau\mu}} < 0.053}     
\newcommand {\KAPPAMlim}  {-0.014 < \kappa < 0.016}     %
\newcommand {\KAPPAElim}  {|\tilde{\kappa}| < 0.26}     %

\begin{document}
%
%
%
\draft
\title{Anomalous charged current couplings of the tau and implications  \\
        for tau compositeness and two-Higgs doublet models}
%
%
\author{Maria-Teresa Dova}
\address{Universidad Nacional de La Plata, La Plata, Argentina}
\author{John Swain and Lucas Taylor}
\address{Department of Physics, Northeastern University, Boston, USA}
\date{\today}
\maketitle
\begin{abstract}
The leptonic branching fractions of the tau lepton are sensitive
to anomalous charged current interactions.
We use recent experimental measurements to determine the weak charged 
current magnetic and electric dipole moments and the Michel parameter
$\eta$ with unprecedented precision.
These results are then used to constrain the tau compositeness scale 
and the allowed parameter space for Higgs doublet models.
\end{abstract}
\pacs{%
12.60.Cn, 
12.60.Fr, 
13.35.Dx, 
14.60.Fg, 
14.80.Cp
}
%
%
\section{Introduction}
The tau lepton in the Standard Model is an exact duplicate of
the electron and muon, apart from its greater mass and
separately conserved quantum number.
Its charged current interactions are expected to be mediated by
the $W$ boson with pure $V\!-\!A$ coupling.
In this paper we present constraints on anomalous charged current 
couplings of the $\tau$ derived from an analysis of the branching 
fractions for 
$\tau^-\rightarrow\mathrm{e}^-\bar{\nu}_{\mathrm{e}}\nu_\tau$ and
$\tau^-\rightarrow\mu^-\bar{\nu}_\mu\nu_\tau$, where charge-conjugate
decays are implied.

In particular, we consider derivative terms in the Hamiltonian which 
describe anomalous weak charged current magnetic and 
electric dipole couplings~\cite{RIZZO97A,CHIZHOV96A} and deviations 
from the $V\!-\!A$ structure of the charged current, to which the 
Michel parameter $\eta$ is sensitive~\cite{MICHEL}. 
The results for the $\eta$ parameter are used to constrain
extensions of the Standard Model which contain more than one 
Higgs doublet and hence charged Higgs bosons.
%
%
\section{Effects of anomalous couplings}
The theoretical predictions for the branching fractions $\Bl$ for the 
decay $\tau^-\rightarrow\ell^-\bar{\nu}_{\ell}\nu_\tau (X_{\mathrm{EM}})$, with
$\ell^-=\mathrm{e}^-, \mu^-$ and $X_{\mathrm{EM}} = \gamma,~\gamma\gamma,~e^+e^-,\ldots$, 
are given by:
\begin{eqnarray}
  \Bl^{\mathrm{theory}}  
              & = & \left(\frac {\GF^2 m_\tau^5}{192\pi^3}\right)\tau_\tau                                         
                    \left( 1 -  8x - 12 x^2{\mathrm{ln}}x + 8 x^3 - x^4\right)                                  \nonumber \\
              &   & \times  \left[ \left( 
                                          1 - \frac{\alpha(m_\tau)}{2\pi} 
                                          \left( 
                                                 \pi^2 - \frac{25}{4} 
                                          \right) 
                            \right) \right.                                                                              \nonumber \\       
              &   & \mbox{~~~~~~}\times \left. \left( 
                                 1 + \frac{3}{5} \frac{m_\tau^2}{\mW^2} - 2 \frac{m_\ell^2}{\mW^2}  
                           \right) \right]                              
                    \left[ 1 + \Delta_\ell \right],
\label{equ:blept}
\end{eqnarray}
where 
$\GF    = \GFMUval$ is the Fermi constant~\cite{PDG96SHORT};
$\tau_\tau = \TAUTval$ is the tau lifetime~\cite{LI97A};     
$m_\tau = \MTval$~\cite{MTAUBESNEW} is the tau mass;
and 
$x=m_\ell^2/m_\tau^2$.
The first term in brackets allows for radiative 
corrections\cite{BERMAN58A,KINOSHITA59A,SIRLIN78A,MARCIANO88A}, 
where $\alpha(m_\tau)\simeq 1/133.3$ is the QED coupling constant~\cite{MARCIANO88A} 
and $\mW = 80.400 \pm 0.075$\,GeV is the $W$ mass~\cite{PIC97}.
The second term in brackets describes the effects of new physics 
where the various $\Delta_\ell$ we consider are defined below.
The sensitivity of these branching ratios to a non-zero neutrino 
mass and mixing with a heavy fourth generation neutrino
has been considered elsewhere~\cite{MNUTAU,MASS_MIX_UPDATE}.                       

The effects of anomalous weak charged current dipole moment 
couplings at the $\tau\nu_\tau W$ vertex are 
described by the effective Lagrangian
\begin{eqnarray}
{\cal{L}}_{\tau\nu W} & = & \frac{g}{\sqrt{2}}\bar{\tau} 
                            \left[ 
                            \gamma_\mu +
                            \frac{i}{2m_\tau}
                            \sigma_{\mu\nu}q^\nu(\kappa_\tau-i\tilde{\kappa}\gamma_5)
                            \right]
                      P_L \nu_\tau W^\mu \nonumber \\
                      & & + ({\mathrm{Hermitian\ conjugate}}),
\end{eqnarray}
where $P_L$ is the left-handed projection 
operator and the parameters $\kappa$ and 
${\tilde{\kappa}}$ are the (CP-conserving) magnetic and (CP-violating) electric dipole form 
factors respectively~\cite{RIZZO97A}.
They are the charged current analogues of the weak neutral current dipole 
moments, measured using $Z\rightarrow\tau^+\tau^-$ events~\cite{PICH97A}, 
and the electromagnetic dipole moments, measured using 
$Z\rightarrow\tau^+\tau^-\gamma$ events~\cite{TAU96_L3,BIEBEL96A,TTGNUCPHYSB}.
In conjunction with Eq.~\ref{equ:blept}, the effects of non-zero 
values of $\kappa$ and ${\tilde{\kappa}}$ on the tau leptonic 
branching fractions may be described by~\cite{RIZZO97A}
\begin{eqnarray}
  \Delta_\ell^{\kappa}         & = &  {\kappa}/{2} + {\kappa^2}/{10};     \\         
  \Delta_\ell^{\tilde{\kappa}} & = &  {\tilde{\kappa}^2}/{10}.            
 \label{equ:deltalk}
\end{eqnarray}
The dependence of the tau leptonic branching ratios on $\eta$ is given,
in conjunction with Eq.~\ref{equ:blept}, by~\cite{STAHL94A}
\begin{eqnarray}
  \Delta_\ell^{\eta} & = &  4{\eta_{\tau\ell}} {\sqrt{x}},                  
\label{equ:deltaln}
\end{eqnarray}
where the subscripts on $\eta$ denote the initial and final 
state charged leptons.
%
Both leptonic tau decay modes probe the charged current couplings of 
the transverse $W$, and are sensitive to $\kappa$ and ${\tilde{\kappa}}$.
In contrast, only the $\tau^-\rightarrow\mu^-\bar{\nu}_\mu\nu_\tau$ channel 
is sensitive to $\eta$ due to a relative suppression factor of $m_e/m_\mu$ for 
the $\tau^-\rightarrow\mathrm{e}^-\bar{\nu}_{\mathrm{e}}\nu_\tau$ channel.
Semi-leptonic tau branching fractions are not considered here since 
they are insensitive to $\kappa$,  ${\tilde{\kappa}}$, and $\eta$.

\section{Results}
We use the recently updated world average values for 
the measured tau branching fractions~\cite{LI97A}:
$\Be = \BRTEval$ and 
$\Bm = \BRTMval$.
Substituting in Eq.~\ref{equ:blept} for 
these and the other measured quantities we obtain
$\Delta_e   = \DELTAEval$ and 
$\Delta_\mu = \DELTAMval$ 
where the errors include the effects of the uncertainties on all 
the measured quantities appearing in Eq.~\ref{equ:blept}.
These results are consistent with zero which, assuming that there are 
no fortuitous cancellations, indicates the absence of anomalous 
effects within the experimental precision.

We therefore proceed to derive constraints on $\kappa$, $\tilde{\kappa}$, and 
${\eta_{\tau\mu}}$ from a combined likelihood fit to both tau decay channels. 
The likelihood is constructed numerically following the 
procedure of Ref.~\cite{NIM_LIKELIHOOD_PAPER} by randomly 
sampling all the quantities used according to their errors, 
conservatively assuming for each parameter that the other 
two parameters are zero.


We determine $\kappa = {\KAPPAMval}$, where the errors correspond 
to one standard deviation, and constrain it to the 
range ${\KAPPAMlim}$ at the 95\% confidence level (C.L.).
This result improves on the 95\% C.L. constraint of 
$|\kappa| < 0.0283$ determined by Rizzo~\cite{RIZZO97A}. 

We determine $\tilde\kappa = {\KAPPAEval}$ and constrain it to the 
range ${\KAPPAElim}$ at the 95\% C.L.
Our constraint, which is the first on this quantity, is considerably 
less stringent than that on $\kappa$ due to the lack of linear terms.
This also means that the results for $\tilde\kappa$
are symmetric by construction.
Were $\tilde\kappa$ to differ significantly from zero, then the likelihood
distribution would have two distinct peaks either side of zero. 
Such structure was not, however, observed.
The decay $W\rightarrow \tau\nu$ is also sensitive to charged current 
dipole terms but, given that the energy scale is $\mW$, the interpretation 
in terms of the static properties $\kappa$ and $\tilde\kappa$ is  
less clear.

We determine ${\eta_{\tau\mu}} = {\ETAval}$ and constrain it to the
range ${\ETAlim}$ at the 95\% C.L.
The uncertainty on our measurement of ${\eta_{\tau\mu}}$ 
is significantly smaller than that obtained by Stahl using the same 
technique $({\eta_{\tau\mu}} = 0.01 \pm 0.05)$~\cite{STAHL94A} 
and more recent determinations using the shape of 
momentum spectra of muons from $\tau$ decays 
$({\eta_{\tau\mu}} = -0.04 \pm 0.20)$~\cite{PICH97A}.
\section{Discussion}
%
%
Derivative couplings necessarily involve the introduction of a length
or mass scale. 
Anomalous magnetic moments due to compositeness are expected to be
of order $m_\tau/\Lambda$ where $\Lambda$ is the compositeness 
scale~\cite{BRODSKY80A}.
We can then interpret the 95\% confidence level on $\kappa$, 
the quantity for which we have a more stringent bound, as a statement
that the $\tau$ appears to be a point-like Dirac particle up to an energy
scale of $\Lambda \approx m_\tau/0.016 = 110$\,GeV. 
These results are comparable to those obtained from anomalous weak 
neutral current couplings~\cite{PICH97A} and more stringent than those 
obtained for anomalous electromagnetic couplings~\cite{TAU96_L3}.

%
Many extensions of the Standard Model, such as Supersymmetry (SUSY),
involve an extended Higgs sector with more than one Higgs doublet. 
Such models contain charged Higgs bosons which contribute 
to the weak charged current with couplings which depend on the fermion masses.
Of all the Michel parameters, ${\eta_{\tau\mu}}$ is especially
sensitive to the exchange of a charged Higgs.
Following Stahl~\cite{STAHL94A}, ${\eta_{\tau\mu}}$ can be written as 
\begin{equation}
{\eta_{\tau\mu}} = -\left( \frac{m_\tau m_\mu}{2} \right)
        \left( \frac{\tan\beta}{m_H} \right)^2
\end{equation}
where $\tan\beta$ is the ratio of vacuum expectation values of the
two Higgs fields, and $m_H$ is the mass of the charged Higgs.
This expression applies to type II extended Higgs sector models 
in which the up-type quarks get their masses from one doublet and 
the down-type quarks get their masses from the other. 

We determine the one-sided constraint ${\eta_{\tau\mu}} > -0.0186$ 
at the 95\% C.L. which rules out the region
    $m_H < (1.86 \tan\beta) \,{\mathrm{GeV}}$ at the 95\% C.L.
%
as shown in Fig.~\ref{fig:chhiggs}.
An almost identical constraint on the high $\tan\beta$ region of 
type II models may be obtained from the process 
$B\rightarrow\tau\nu$~\cite{HOU93A}.
The most stringent constraint, from the L3 experiment,
rules out the region
    $m_H < (2.09 \tan\beta)\,{\mathrm{GeV}}$ at the 95\% C.L.~\cite{L3B2TAU}.
Within the specific framework of the minimal supersymmetric standard 
model, the process $B\rightarrow\tau\nu X$ rules out the region 
$m_H < (2.33 \tan\beta)\,{\mathrm{GeV}}$ at the 
95\% C.L.~\cite{COARASA97A}.
This limit, however, depends on the value of the Higgsino mixing 
parameter $\mu$ and can be evaded completely for $\mu>0$. 
The non-observation of proton decay also tends to rule out 
the large $\tan\beta$ region but these constraints are 
particularly model-dependent.
The very low $\tan\beta$ region is ruled out by measurements of 
the partial width $\Gamma (Z\rightarrow b\bar{b})$. 
For type II models the approximate region excluded is 
$\tan\beta < 0.7$ at the $2.5\sigma$ C.L. for 
any value of $M_H$~\cite{GRANT95A}.  
Complementary bounds for the full $\tan\beta$ region are derived 
from the CLEO measurement of 
$BR(b\rightarrow s\gamma) = (2.32\pm0.57\pm 0.35)\times 10^{-4}$  
which rules out, for type II models, the region 
$M_H < 244 + 63/{(\tan\beta)}^{1.3}$~\cite{CLEO_BTOSGAMMA}.
This constraint can, however, be circumvented in SUSY models
where other particles in the loops can cancel out the effect
of the charged Higgs.
Direct searches at LEP II exclude the region $m_H < 54.5$\,GeV for 
all values of $\tan\beta$~\cite{DELPHIMH}.
The CDF search for charged Higgs bosons in the process
$t\rightarrow b H^+$ rules out the region of low $m_H$ and 
high $\tan\beta$~\cite{ABE97A}.

The 95\% C.L. constraints in the $m_H$ {\em{vs.}} $\tan\beta$ plane, 
from this and other analyses, are shown in Fig.~\ref{fig:chhiggs}.
\vbox{%
\begin{figure}
   \mbox{\epsfig{file=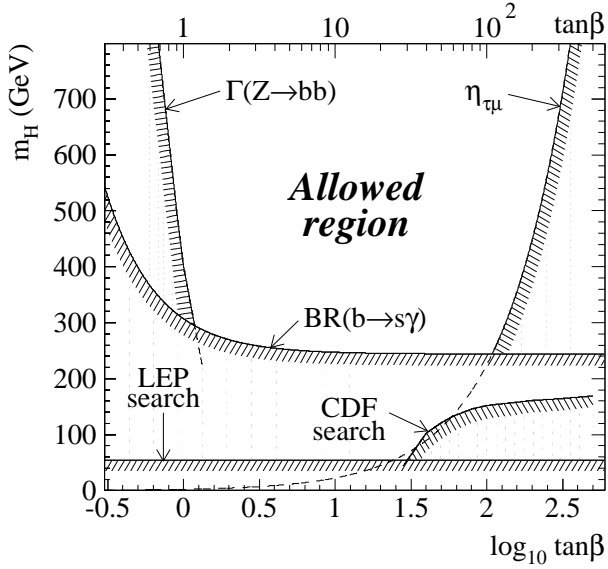,width=0.45\textwidth,clip=}}
  \caption{Constraints on $m_H$ as a function of $\tan\beta$ at the 95\% C.L.,
           from this analysis of $\eta_{\tau\mu}$ and the other analyses 
           described in the text.}\label{fig:chhiggs}
\end{figure}
}%
\section{Summary}
From an analysis of tau leptonic branching fractions we determine
\begin{eqnarray}
\kappa           & = & {\KAPPAMval};          \\
\tilde\kappa     & = &  {\KAPPAEval};         \\
{\eta_{\tau\mu}} & = &  {\ETAval}.            
\end{eqnarray}
Each of these results is the most precise determination to date.
The result for $\kappa$ indicates that the tau is point-like up
to an energy scale of approximately 110\,GeV.
The result for ${\eta_{\tau\mu}}$ constrains the charged Higgs 
of type II two-Higgs doublet models, such that the region 
\begin{equation}
m_H < (1.86 \tan\beta) \,{\mathrm{GeV}}
\end{equation}
is excluded at the 95\% C.L.
\section*{Acknowledgements}
We would like to thank Carlos Garc\'{\i}a Canal for bringing the topic
of derivative couplings in tau decays to our attention,
and for his continuing support.
MTD acknowledges the support of CONICET, Argentina. 
JS and LT would like to thank the Department of Physics,
Universidad Nacional de La Plata for their generous hospitality 
and the National Science Foundation for financial support.
JS gratefully acknowledges the support of the International
Centre for Theoretical Physics, Trieste.

\end{document}